# Supporting Dynamic Agentic Workloads: How Data and Agents Interact

Ioana Giurgiu
*IBM Research Europe*
*Zurich, Switzerland*
igi@zurich.ibm.com

Michael E. Nidd
*IBM Research Europe*
*Zurich, Switzerland*
mni@zurich.ibm.com

*Abstract*—The rise of multi-agent systems powered by large language models (LLMs) and specialized reasoning agents exposes fundamental limitations in today's data management architectures. Traditional databases and data fabrics were designed for static, well-defined workloads, whereas agentic systems exhibit dynamic, context-driven, and collaborative behaviors. Agents continuously decompose tasks, shift attention across modalities, and share intermediate results with peers – producing non-deterministic, multi-modal workloads that strain conventional query optimizers and caching mechanisms.

We propose an Agent-Centric Data Fabric, a unified architecture that rethinks how data systems serve, optimize, coordinate, and learn from agentic workloads. To achieve this we exploit the concepts of attention-guided data retrieval, semantic micro-caching for context-driven agent federations, predictive data prefetching and quorum-based data serving. Together, these mechanisms enable agents to access representative data faster and more efficiently, while reducing redundant queries, data movement, and inference load across systems.

By framing data systems as adaptive collaborators, instead of static executors, we outline new research directions toward behaviorally responsive data infrastructures, where caching, probing, and orchestration jointly enable efficient, context-rich data exchange among dynamic, reasoning-driven agents.

## I. INTRODUCTION

Recent advances in large language models (LLMs) and autonomous reasoning agents are transforming how computation, coordination, and decision-making are organized [1] [2] [3]. In contrast to traditional static workloads, multi-agent systems operate as dynamic, evolving ecosystems where agents collaborate, adapt, and generate data-intensive tasks in real time [4] [5] [6]. These agentic systems are increasingly used for complex analytics, data integration, and scientific workflows, yet the underlying data management infrastructures remain largely unchanged.

Today's databases are designed for predictable, declarative workloads: queries are well-defined, data schemas are stable, and cost-based optimizers can plan execution deterministically [7] [8] [9]. However, agentic workloads are fundamentally different. Agents continuously decompose goals into evolving sub-tasks, refine their data needs as reasoning unfolds, and exchange intermediate results with other agents [10] [11]. This results in non-deterministic, multi-modal, and highly contextual workloads, spanning structured, unstructured, and streaming data sources. Such workloads stress every layer of modern data systems, from query parsing to caching, optimization, and execution.

### A. Limitations of Current Systems

Existing data systems are not designed to accommodate these new patterns of behavior-driven data access. *Static query planning* assumes full query visibility upfront. In contrast, agentic queries emerge progressively as agents reason or collaborate. *Traditional caching mechanisms* [12] [13] index results by syntactic query strings, failing to capture semantic similarity or reuse opportunities across related agent prompts or subgoals. *Engine-centric architectures* isolate computation by modality (e.g., SQL engines, vector stores, streaming processors) [14] [15] [16], forcing agents to manually coordinate data retrieval and reconciliation. *Cost models* are optimized for I/O and CPU usage, not for token costs, inference latency, or energy consumption of machine learning operators [17] [18]. The result is an inefficiency gap: agents often over-query, repeatedly retrieve similar data, or re-compute embeddings and summaries that could have been shared. Meanwhile, databases become overloaded with semantically redundant requests and transient analytical bursts. A new generation of agent-aware data systems is needed – ones that treat data serving and coordination as adaptive, context-sensitive processes rather than static processes.

### B. The Case for an Agent-Centric Data Fabric

*What does it mean to optimize behaviors instead of queries in data systems?* To answer this question, we propose an Agent-Centric Data Fabric, a unified architecture that redefines how data systems interact with dynamic, reasoning-driven agents. In our vision, the data system becomes an active collaborator, capable of anticipating, adapting, and learning from agentic workloads. Rather than passively executing SQL or API calls, the data fabric mediates information flow between agents, data sources, and inference engines through a combination of probing, caching, prefetching and coordination mechanisms.

Specifically, we propose several foundational mechanisms:
1) *Semantic micro-caching* – storing and retrieving data for agents in the same semantic federation by meaning rather than query text, reducing repetitive lookups and enabling knowledge reuse.

2) *Attention-guided data retrieval and prefetching* - predicting and prefetching the most relevant data based on agent intent and task context.
3) *Quorum-based data serving* - enabling shared access to verified, representative data across collaborating agents while minimizing redundant retrieval.

Together, these mechanisms transform data systems from static executors into adaptive collaborators, capable of dynamically serving the evolving data needs of autonomous agents.

Our vision proposes a paradigm shift: from data systems optimized for queries, to data systems optimized for behaviors. In such an ecosystem, agents can efficiently access, share, and refine data without overloading backend systems. Caching becomes predictive and semantic, data movement becomes context-driven, and orchestration layers continuously adapt to workload dynamics.

We first motivate the need for such a shift by highlighting the behavioral dynamic of agentic workloads. Then, we propose an architecture that is able to deal with these dynamics and outline open challenges to realizing this vision.

## II. Behavioral Dynamics of Agentic Workloads

Agentic workloads exhibit runtime variability, self-modifying behavior, and context-dependent coordination, fundamentally challenging traditional data management assumptions. Unlike deterministic workloads where queries, schemas, and execution paths are static, agentic workloads evolve continuously as agents decompose tasks, react to feedback, and share results. These dynamics directly influence how data is discovered, probed, served, and reused across the system.

Consider a multi-agent system deployed by a global logistics company to predict, explain, and mitigate delivery disruptions across its worldwide supply chain. The system comprises the following specialized agents:

1) Orchestrator: monitors all agents, collecting latency, cost, and accuracy metrics; acts as a meta-agent, continuously refining global coordination.
2) Anomaly detection: continuously monitors structured shipment data and live events to identify unusual patterns; triggers downstream analysis when anomalies are detected.
3) Sentiment analysis: analyzes unstructured customer feedback to add qualitative signals.
4) Root cause analysis: fuses anomaly statistics, live events, and sentiment embeddings; correlates structured anomalies with semantic evidence to determine root cause for delivery disruptions.
5) Forecasting: uses identified root causes and historical data to predict future delay trends.
6) Routing optimization: consumes predictions and root cause signals to (re)plan delivery routes.

A natural flow for these agents is shown in Figure 1. They operate over structured, unstructured, and streaming data sources, as shown in Table I.

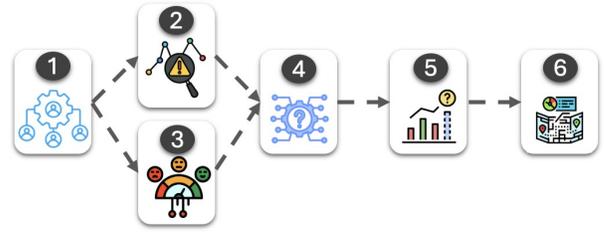

Fig. 1: Workflow for motivating example.

TABLE I: Example data sources.

| Data type | Example source | Example content |
|---|---|---|
| Structured | `shipments_db` (PostgreSQL) | ship_id, region, eta, act_delivery, carrier, status |
| Unstructured | `cust_feedback` (NoSQL / Text) | JSON: *{customer_id, text, timestamp}* |
| Streaming | `live_stream` (Kafka) | *{event_type, region, delay_min, source}* (weather alerts / customs delays) |

### A. Dynamic Task Decomposition and Query Generation

> Agents dynamically refine and recompose their goals during execution, altering query structures and data dependencies at runtime.

Agents are able to modify their reasoning strategy and task graph at runtime. Unlike static workflows, where the set of queries, operators, and dependencies is known at submission time, agentic systems exhibit evolving workloads whose structure depends on intermediate results, user feedback, and contextual cues derived from the environment. Each agent typically begins with a high-level intent, such as 'analyze delivery anomalies'. As agents reason through subtasks, they may generate new subgoals, refine prior plans, or delegate partial results to other agents. This behavior creates non-deterministic query graphs whose topology – number of nodes, order of execution, and operator types – evolves as reasoning unfolds. From a data management perspective, this means query boundaries blur: databases must support potentially increased probing, and adaptive caching as agents shift focus between structured and unstructured data. A SQL aggregation may seamlessly evolve into a hybrid pipeline combining embeddings, sentiment analysis, and predictive inference.

*Back to example:* Initially, the Anomaly detection agent issues a SQL query to compute daily delivery delays by region:

```
SELECT region, AVG(act_delivery-eta) AS avg_delay
FROM shipments
WHERE status='delivered'
GROUP BY region;
```

Upon detecting high delay variance in Southeast Asia, the agent decomposes its task into 3 subtasks:

1) 'find root cause' → route to Root cause analysis agent;

2) 'gauge customer perception' → route to Sentiment analysis agent;
3) 'forecast impact' → route to Forecasting agent.

These new subgoals emerge at runtime – the workflow's shape is not known a priori – and the query graph evolves as new evidence arrives.

*B. Speculative Querying*

Agents speculate, forming probing queries to test which data sources, modalities, or endpoints yield meaningful signals for their task.

When agents lack full schema or data knowledge, they engage in speculative probing, issuing exploratory micro-queries to learn what data exists and its usefulness [19]. Current database engines are optimized either for long-running OLAP queries or high-frequency OLTP queries. Speculative probing falls into neither category, because it creates a high fan-out of tiny queries, performing schema-level operations (i.e., schema sampling, introspection queries, lightweight metadata scans), with low-temporal locality (i.e., each probe touches new columns or tables). As a result, cost optimizers cannot meaningfully predict or amortize cost across random probes. Additionally, metadata locks (e.g., in PostgreSQLs `pg_class`) become contention points under hundreds of concurrent probes.

To accommodate this behavior, data systems need to optimize probing, support rapid, low-overhead metadata access and amortize the cost of thousands of concurrent probes.

*Back to example:* The Anomaly detection agent is tasked with finding unusual delivery delays, but it has only partial knowledge of a logistics database, namely it knows the database contains tables about shipments, deliveries, and customers, but not their exact names or schemas. The agent starts by issuing schema-level probes to gather metadata, by searching for tables semantically related to its goal (e.g., '%delivery%', '%delay%'). It learns that for instance tables `shipments` in `shipments_db` is relevant. Then it proceeds to probe around the data contained in the columns and to understand their meaning:

```
SELECT act_delivery, eta FROM shipments LIMIT 5;
```

From a few rows, it infers that `act_delivery` and `eta` are timestamps, thus suggesting that delays could be computed from their difference. Based on this learned structure, the agent hypothesizes that `delivery_delay = act_delivery - eta`. It then formulates a speculative query:

```
SELECT region, AVG(act_delivery -eta) AS avg_delay
FROM shipments
GROUP BY region;
```

The agent runs this query and tests whether results exhibit a sign of anomaly. If variance is low, the agent revises its hypothesis – perhaps another table better captures the target variable. Such a process is highly iterative and imposes significant load on the underlying data system, particularly when faced with concurrent requests.

*C. Context-Driven Data Access*

Agents reason over evolving context, maintaining short-term and shared memory that shapes subsequent queries.

Unlike stateless query clients, agents reason over context – both their own (short-term memory) and that of the larger system (shared long-term state). Context becomes a first-class data modality. As a result, agents reuse prior reasoning fragments, inferred schemas or computed embeddings, and coordinate through shared memory to guide new data access paths.

This demands context-aware indexing and caching layers that exploit semantic similarity, provenance, and freshness to prevent redundant computation. Data systems must support semantic reuse policies, where context embeddings determine cache hits or prefetch opportunities.

*Back to example:* The Root cause analysis agent will use structured, streaming and unstructured information to detect a root cause for delays. Since the Sentiment analysis agent already computed embeddings reflecting customer reactions, it could reuse these embeddings to semantically connect 'customs issues' in text data with structured `event_type='Customs Delay'`.

*D. Collaboration Among Agents*

Agents exchange intermediate results – structured data, embeddings, or textual summaries – that affect the behavior of other agents.

In multi-agent data ecosystems, agents rarely operate in isolation. Instead, they collaborate through intermediate outputs that are dynamically exchanged during task execution. This collaboration introduces a new dimension of data flow semantics, where intermediate results produced by one agent become inputs to another. Unlike traditional distributed query plans, these communication patterns are not precompiled but evolve at runtime, shaped by agent reasoning, feedback, or environmental changes.

Agents also vary in their reasoning modes and modalities: one may output structured tables, another embeddings, and a third natural language summaries. Hence, collaboration must bridge heterogeneous data representations and modalities, requiring systems to reason about semantic equivalence, not just structural compatibility. For instance, two agents may exchange information referring to 'server latency' and 'response time', which need to be resolved semantically before further processing.

*Back to example:* The Anomaly detection, Root cause analysis and Routing optimization agents communicate via a pub/sub bus. The Anomaly detection agent publishes a structured summary:

```
{region: "Singapore", anomaly_score: 0.92,
```

```
    correlated_keywords: ["customs"]}.
```

The Root cause analysis and Routing optimization agents subscribe to these messages. The Root cause analysis agent fetches additional multimodal evidence, and returns a synthesized explanation to the Orchestration agent, whereas the Routing optimization agent takes input from both other agents, reoptimizes delivery routes and publishes the updated route plans:

```
{"region": "Southeast Asia", "new_route": "Singapore
    >> Kuala Lumpur >> Destination"}
```

### E. Agent Self-Evaluation

> Agents introspect on their own performance, learning when to adjust strategies, caching, or resource use.

Unlike traditional databases or machine learning models, which follow static execution plans, agents continuously monitor their own performance, by tracking latency, token cost, and accuracy, and use these signals to adjust strategies, caching, or model selection. This creates closed feedback loops both at the agent and system level: an agent may learn to prefetch embeddings for frequent queries, while an orchestrator aggregates feedback to adjust global scheduling and caching policies.

Supporting such feedback-driven adaptation requires introspective telemetry (e.g., cache hit rates, inference costs) and policy learning mechanisms. Over time, systems can evolve into self-tuning data fabrics that learn from agents' operational histories.

*Back to example:* The Orchestrator agent continuously monitors latency, cost, and model accuracy across all agents and collects telemetry, such as:

```
Agent      | Latency | Model      | Acc  | Token Cost
---------------------------------------------------------
Sentiment  | 220 ms  | DistilBERT | 0.87 | $0.01
Root Cause | 480 ms  | Llama2     | 0.91 | $0.04
```

It will use this type of feedback to update cost models, adjust model-switch thresholds and recommend potential caching strategies. Over time, it can even learn policies such as: 'Use lightweight models for frequent, low-risk regions; prioritize high-accuracy inference for anomalies with high severity'.

## III. ARCHITECTURE OF THE AGENT-CENTRIC DATA SYSTEM

Our envisioned architecture for the agent-centric data system is shown in Figure 2.

Initially, an agent issues a high-level multimodal task (e.g., 'analyze product defects using QA logs, sensor streams, and customer reviews'). The **Agent layer** activates a context federation a cluster of collaborating agents that share embeddings, prior reasoning traces, and micro-caches. The **Attention-guided data retrieval** ① module computes an attention distribution over possible data sources, based on query semantics, past behavior, and context embeddings. The attention weights determine which data partitions are most salient for early exploration. The agent's query and attention signal are passed to the Orchestration layer.

In the **Orchestration Layer**, the **Attention-guided router** ③ issues lightweight speculative probes to candidate engines based on saliency weighting (i.e., using attention scores to assign priorities). These requests are used to estimate relevance, latency, and cost before committing to full retrieval. The **Predictive prefetcher** ④ preloads likely next-access data (e.g., relevant embeddings or cached results) into caches. Local **Micro-caches** ② (in the Agent layer) hold immediate, task-specific results. The shared semantic caches store long-term reusable embeddings or results across agents.

The **Optimizer** ⑤ continuously learns from KPIs collected by the **Monitoring** component ⑩ (in the Execution layer) and agent behavior to refine routing, caching, and cost models across heterogeneous backends. It provides optimized plans for a query to ③, and adjusts routing, caching and quorum thresholds. ③ takes the optimized plans, compiles subplans into engine-specific executable queries or code fragments, and routes them accordingly to the Execution layer. In parallel, the **Cross-agent cache manager** ⑥ continuously monitors embeddings of active tasks to detect semantic overlap between agents in ② and the **Shared semantic cache** ⑦. When overlap is detected cached embeddings or partial results are merged, duplicate probes are suppressed, and relevance scores are shared across federations. The goal is to reduce redundant backend queries and foster cooperative data reuse, by essentially creating a distributed, semantic memory shared among agents.

Once engines in the **Engine fabric** ⑨ begin to process their respective subqueries in the **Execution layer**, results begin to stream in. This is where the **Quorum-based serving** component ⑧ steps in. It enables confidence-driven data serving: instead of waiting for all backends to respond, it aggregates partial results until a semantic quorum is reached and then serves early results to the requesting agent or federation.

Finally, the orchestration layer merges results from multiple engines into a coherent multimodal output (e.g., a table enriched with textual insights or vector matches). Agents receive results asynchronously and dependent agents in the workflow (e.g., Root Cause Analysis after Anomaly Detection) are triggered by message-passing events.

Next, we dive deeper into each of the three layers.

### A. Agent Layer

The Agent Layer mediates between high-level intent and concrete data access by incorporating two complementary mechanisms: attention-guided data retrieval, which focuses queries toward the most relevant sources, and micro-caches, which locally retain frequently accessed context for efficiency. **Attention-guided data retrieval**. Traditional query systems assume a complete, explicit query, but agentic systems often operate under partial knowledge. An agent knows roughly what it needs but not where it resides or how best to express it. Attention mechanisms – introduced in transformer architectures [20] – provide a natural formalism for expressing this uncertainty: they encode relevance distributions over data sources or features rather than discrete access decisions.

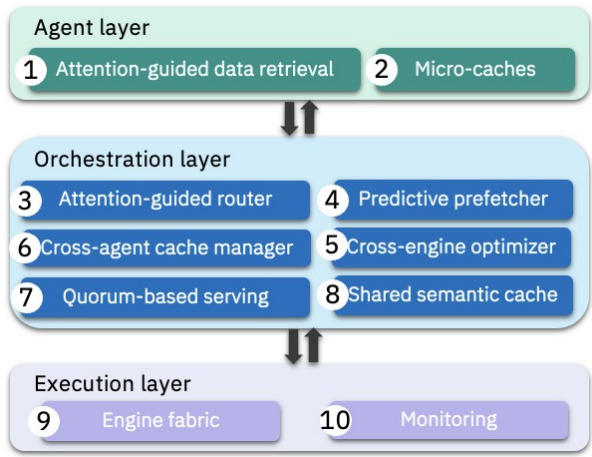

Fig. 2: Architecture for the proposed agent-centric data system.

Conceptually, each agent maintains an attention vector representing its current focus, derived from task embeddings (e.g., encoded prompt, intent, or user goal), context embeddings (e.g., previous results, feedback, conversation history) and data embeddings (e.g., table summaries, schema vectors, metadata). These vectors interact through similarity metrics to yield attention weights over data partitions, tables, or modalities:

$$w_i = \frac{exp(sim(q, d_i)/\vartheta)}{\sum_j exp(sim(q, d_j)/\vartheta)} \quad (1)$$

where q is the agent's query embedding, $d_i$ is the embedding of a data partition or source (precomputed via schema summarization, metadata embeddings, or learned representations), and $\vartheta$ controls selectivity. The system interprets these weights as soft routing probabilities, which act as a guide for incremental probing and prefetching operations before issuing full queries. This attention-based mechanism has the potential to reduce data scanning and tail latency.

We outline the mechanics of the attention-guided data retrieval briefly. Similarity scores between an agent's context vector and data summaries are computed. Instead of querying all candidate sources, the agent issues speculative probes proportional to attention weights. These lightweight metadata requests allow to quickly assess relevance. As feedback is received either from probes or collaborating agents, the attention distribution sharpens, enabling the system to promote with high confidence data sources for full data retrieval. Finally, KPIs about query success, latency and accuracy are used to update future focus distributions.

**Shared micro-caches**. Because many agents operate on overlapping data scopes and intent regions, repeated access to the same embeddings, intermediate results, or sub-plans can severely strain backend systems.

To mitigate redundant access, we propose that agents within a context federation (i.e., working on similar tasks) share a micro-cache, namely a lightweight, semantically indexed repository. Each cache stores small, high-utility data fragments such as embeddings of previously accessed records or entities, partial query results or vector search hits, metadata sketches, histograms from probes, or model inference outputs (e.g., classification labels or summaries).

The mechanics of the micro-caches are as follows. When an agent completes a query or probe, they are encoded as embeddings and inserted into the micro-cache. Before launching a new query, an agent checks the cache for close matches (e.g., cosine similarity above a learned threshold). Matched entries can be reused directly or adapted. Cache retention policies are guided by reuse likelihood and attention frequency. The intuition behind the latter is that if a region of data (e.g., schema fragment, or stream window) repeatedly attracts high attention weights, it is a strong candidate for prefetching or retaining in the cache. Conversely, data regions with low or decaying attention frequency are evicted or compressed.

### B. Orchestration Layer

The Orchestration Layer mediates between dynamic agent queries and heterogeneous data engines. It monitors evolving agent intents, predicts data needs, probes selectively, and orchestrates data movement and computation across diverse engines.

Its core design principles are as follows: (1) dynamic adaptivity at runtime; (2) cross-modality awareness; (3) collaborative optimization by identifying overlapping intents to merge or synchronize queries and minimize redundant data retrieval; (4) progressive refinement of routing and caching strategies based on KPIs and attention patterns.

We envision the orchestration layer to be composed of the following key components.

**Cross-engine optimizer**. The Optimizer serves as the adaptive learning core of the architecture, continuously refining how data is routed, cached, and served across heterogeneous backends. Unlike traditional query optimizers that operate once per query, the optimizer functions as a feedback-driven control layer that learns from KPIs, agent behavior, and execution outcomes. It updates cost and performance models for diverse operators, ranging from SQL joins to vector similarity searches and model inference, and adjusts routing, caching, and quorum thresholds accordingly. By monitoring query latency, confidence, and reuse patterns, it tunes probe aggressiveness (i.e., how many speculative sources to contact, the depth of vector or index traversal, or the confidence threshold for initiating a probe), cache lifetimes, and modality priorities, effectively guiding the Attention-guided router, Cross-agent cache manager, Predictive prefetcher and Quorum-based serving components through learned policies rather than static heuristics.

To enable this continuous learning process, the Execution layer streams back KPIs (latency, throughput, cache hits, inference cost) to the optimizer. This feedback loop (e.g., which routing decisions led to low latency or high confidence, which backends performed poorly under load, how often probes yielded useful data, how often cached results avoided re-querying, etc.) forms the training data for its next learning iteration. It also allows for adaptive optimization to take place.
**Attention-guided router**. The router is responsible for transforming high-level agentic intents into efficient, adaptive query execution plans. It continuously integrates attention signals, runtime feedback, and optimizer directives to decide what data to access, where to execute it, and how much computation to invest. Building upon the attention mechanisms already active in the Agent layer, the router learns a relevance-weighted mapping between the agent's query embedding $q$ and candidate data partitions or modalities $d_i$ (Eq. 1). Using these attention weights, the router performs two complementary roles:

1) *speculative probing* – the router initiates lightweight probes across high-weight partitions or modalities. Based on assessed data quality and expected cost, irrelevant or expensive paths can be pruned early and high focus is placed on data sources with high semantic relevance and acceptable cost.
2) *execution of optimized plans* – the router becomes the operational driver of the optimizer. It routes subqueries and operators to the most suitable backends and can redirect subqueries when re-optimization occurs.

The router essentially balances exploration (via probing) and exploitation (via optimized execution). It ensures that agents access only salient data while maintaining adaptability to evolving workloads.

**Predictive prefetcher**. The prefetcher anticipates future data needs by learning access patterns and task sequences from past agent interactions. Using information from the router and optimizer, it identifies data regions that agents are expected to query next with high likelihood. Prefetching occurs both speculatively (triggered by probes assessing relevance and cost) and cooperatively, where multiple agents' intents overlap and prefetches are shared via the semantic cache. Prefetched data is first stored in local micro-caches for low-latency reuse, then promoted to shared semantic caches when confirmed useful across agents.

The prefetching engine essentially learns temporal correlations between agent goals and subsequent data accesses. Ideologically, it is similar to sequence modeling in workload prediction [21] [22] [23].

**Cross-agent cache manager**. The cache manager monitors the embedding space of active tasks across agents and federations. Each agent's current intent, intermediate output, or query vector is compared against others using semantic similarity metrics. When a significant overlap is detected, thus indicating that multiple agents are reasoning about related data regions, entities, or modalities, it intervenes to coordinate reuse rather than recomputation, by merging results into the **Shared semantic cache**.

Additionally, when semantic similarity across agents' probes exceeds a configurable threshold, redundant queries are suppressed before reaching the execution layer. The Cache manager either redirects the agent to an existing cache entry or delays redundant probe issuance until the active one completes, significantly reducing backend I/O and inference load.
**Quorum-based serving**. In a traditional database, a query planner produces a fixed plan, sends it to a single engine, and waits for full results before returning an answer. For multi-agent workloads, this approach is ineffective because agents issue probabilistic, approximate, or multi-engine queries. In addition, some backends (e.g., vector stores, inference servers) are slow, unreliable, or non-deterministic, and results are often aggregated from heterogeneous modalities.

We approach this by allowing partial, confidence-based completion of a query rather than waiting for every system to respond. We formulate that a *semantic quorum* is reached when enough information has been gathered to make a confident decision, namely:

$$conf = f(coverage, diversity, agreement) > \vartheta \quad (2)$$

where coverage = how many data partitions have responded, diversity = how representative or independent the responses are (e.g., embeddings from different sources, textual vs. numerical evidence), agreement = how consistent the responses are (e.g., similar vectors, overlapping entities, or corroborating values).

Once confidence exceeds threshold $\vartheta$, the coordinator can return early, enabling lower latency by enabling the system to return good-enough answers faster when perfect completeness is unnecessary, reduce system load, and degrade gracefully under missing or slow components. Additionally, it introduces a structured way to measure agreement across heterogeneous modalities.

*C. Execution layer*

This layer unifies heterogeneous data and computation backends via an **Engine fabric**. It integrates relational databases, vector stores, stream processors, and inference servers. An optimized plan generated from the Orchestration layer can be executed across engines by the Execution layer. Instead of using shared intermediate representations such as Substrait [24] which require alignment from all engines and significant extensions to support different modalities, we propose a hybrid model that combines a minimal representation with on-demand compilation and open data sharing protocols. the minimal representation is a lightweight, system-agnostic abstraction for describing computational intent, namely what operations are performed, in what order, and with which semantic guarantees, without committing to any particular execution engine or data modality. The on-demand compilation transforms subplans into native execution units for each target backend. Data sharing protocols standardize how datasets, metadata, and lineage are exposed, versioned, and transmitted between engines.

Complementing the Engine fabric, the **Monitoring** component allows the Optimizer to continuously observe and learn from execution. It collects runtime metrics – including latency, throughput, accuracy, energy use, and cost – from all engines and agents. These are used to refine cost models, adjust query routing policies, and tune caching and probing strategies based on observed workload dynamics.

## IV. OPEN CHALLENGES

### A. Unified cost modeling across engines and modalities

A foundational yet unresolved problem in agent-centric systems is the development of cost models that span heterogeneous data types, execution engines, and operator semantics. In agentic systems, each operator may consume different resources (e.g., GPU time, token budgets, network bandwidth) and exhibit stochastic latency or quality characteristics. Moreover, agents often optimize not only for runtime but also for token-cost efficiency, model accuracy, and energy consumption, creating a multi-objective optimization space typically absent from classical systems.

Open research directions include designing composable cost abstractions that unify deterministic and probabilistic operators under a common metric space. Work on cross-platform optimizers such as RHEEM/Apache Wayang [14] and Musketeer [16] illustrates early attempts to estimate cost across heterogeneous engines, yet they remain tied to relational and dataflow operators. Similarly, adaptive and learned optimization techniques [25] [18] [17] demonstrate that feedback-driven cost learning can outperform static heuristics, but extending these ideas to multimodal workloads and stochastic inference remains open. Such a model must represent heterogeneous costs – CPU/GPU utilization, tokenized inference costs, I/O latency, and network transfer – while remaining expressive enough for cross-engine planning.

A related challenge is cost calibration under uncertainty: how to predict effective cost when operator latency distributions or model inference times vary with input semantics, as observed in approximate nearest-neighbor (ANN) search benchmarks [26] and large-model serving systems such as vLLM [27]. Another open question concerns costutility pairing – how to incorporate non-monetary outcomes such as accuracy, confidence, or semantic coverage into cost-based decision frameworks, enabling the optimizer to trade precision for timeliness or price for fidelity. Finally, maintaining a learned cost model that adapts across modalities and workloads (e.g., structured SQL operators, approximate ANN indexes, or transformer inference) requires KPI-driven feedback loops and transfer-learning mechanisms that preserve stability while adapting to drift.

Developing such a unified, multimodal cost framework would underpin nearly every adaptive component of the proposed architecture – from attention weighting and prefetch decisions to cross-engine routing and optimization.

### B. Cost- and context-aware attention mapping

The attention-guided router depends on computing a relevance distribution that reflects not only semantic similarity but also system-cost factors (e.g., inference latency, data volume, cost per token). While existing attention mechanisms [20] focus purely on relevance, future work must augment attention models with cost features. Possible approaches include multi-armed-bandit or reinforcement-learning frameworks (inspired by Cuttlefish [25]) where each source $d_i$ is associated with a costutility reward, and attention weights evolve by maximizing utility (relevance/cost trade-off). Further, context drift triggered by agents iteratively reformulating tasks could be addressed by online adaptation of attention weights – an avenue where adaptive query processing literature [28] [29] offers insights into feedback-driven optimization loops.

### C. Probing and prefetching under uncertainty

In our vision, probing and prefetching are used as complementary strategies to balance information gain, latency, and system load. Techniques like approximate query processing [30] [31] [32] could offer inspiration. However, they cannot be applied directly because agents' intents evolve dynamically, leading to changing relevance distributions and non-stationary access probabilities.

Formulating probe utility functions that jointly consider information gain, latency, and resource cost therefore remains an open research problem. Additionally, an effective predictive prefetching means that given a current intent, the system must estimate a probability distribution over future data access and optimize prefetch timing and fidelity. Such a problem could be modeled as a Markov decision process [33] or with look-ahead information-passing techniques [34] [20]. Open directions include learning accurate access-sequence models (e.g., via LSTM or transformer architectures) across asynchronous agent workflows, and coordinating prefetching policies to avoid prefetch storms that overload shared resources.

### D. Cost-effective and consistent semantic caching

Recent advances in semantic caching for LLM and multimodal inference [35] [36] [37] [38] [39] move beyond exact-match caching toward embedding-based reuse of semantically similar prompts or intermediate representations. These systems directly inform the design of the proposed micro-caching layer, where agent-local reuse depends on intent and task semantics. For instance, vCache's [35] verified reuse introduces formal error bounds for cached responses, while KVShare [36] and SentenceKV [37] demonstrate finer-grained reuse of key (value or sentence) level representations to reduce redundant computation and latency within individual agents. Together, these approaches illustrate how the micro-caches and the shared cache proposed in our architecture could integrate confidence-aware and semantically adaptive reuse without requiring full query re-execution.

However, these systems largely operate within the scope of a single model or user session and thus do not yet address the challenges of shared semantic caches envisioned in our architecture. In a shared cache spanning multiple autonomous agents and modalities, new research questions arise that current LLM-oriented caching work leaves open. First, cross-agent semantic coherence: how to detect and merge semantically overlapping cache entries across heterogeneous agents, embedding models, and modalities without costly full-vector scans remains unsolved. Second, federated consistency and privacy: existing systems lack mechanisms for maintaining freshness or enforcing isolation across tenants, which are critical for preventing stale or unintended reuse in multi-agent federations. Third, adaptive costutility modeling: while vCache and KVShare optimize reuse for accuracy or latency, an agentic data system must balance multi-objective costs – inference price, token consumption, and retrieval latency – when promoting items from micro-cache to shared cache. Finally, semantic conflict resolution remains open: when agents cache partially overlapping or contradictory results, the system must reconcile or version them in provenance-aware ways, possibly integrating trust weighting or quorum-based validation.

*E. Consistency vs. latency in quorum-based serving*

The Quorum-based serving coordinator introduces a powerful yet underexplored idea: allowing partial results to be returned once confidence thresholds are met across diverse modalities. Realizing this capability raises several open research challenges.

A first challenge is how to combine uncertainty estimates from heterogeneous sources such as structured queries, vector joins, and LLM inferences into a unified probabilistic framework. Techniques from ensemble learning [40] and model calibration [41] provide starting points, but extending them to multimodal systems may require modality-specific uncertainty measures (e.g., Bayesian networks for text, variance metrics for vector similarity).

A second challenge concerns adaptive quorum thresholds, where $\vartheta$ should dynamically adjust to workload intent or SLA priority, thus allowing mission-critical queries to demand higher confidence than exploratory tasks. Ideas from incremental and adaptive query processing [42] [43] [44] can be borrowed, but need to be extended to probabilistic operators. Finally, handling late-arriving or revised results remains unresolved: when slower backends or higher-fidelity engines return divergent outcomes, the system must reconcile them through result refinement or revision transparency to agents, ensuring trust and auditability.

*F. Data protocols and minimal IRs*

Research challenges emerge in maintaining the alignment between the minimal IR and data sharing protocols – defining protocol bindings that preserve semantics across modalities, supporting consistent versioning and provenance through shared metadata, and generating executable code that respects both the logical intent of the IR and the physical constraints advertised by each engine. Achieving this balance remains a central open problem in enabling portable yet semantics-preserving multimodal query execution.

*G. Additional open challenges*

While the preceding sections detail core challenges for each architectural component, several cross-cutting issues remain open.

First, cross-component optimization is largely unexplored. The proposed layers of cost modeling, semantic caching, and quorum-based serving interact closely in practice: cache behavior affects quorum thresholds, while quorum outcomes influence cost estimation and attention priorities. Developing joint optimization frameworks that coordinate these components through shared feedback and adaptive control remains a key challenge.

Second, temporal adaptation and drift management across layers require deeper study. Representations underlying caches, attention maps, and cost estimators evolve as agents, data, and models change. Detecting and mitigating representation drift without full recomputation is crucial for maintaining consistency and performance over time.

Third, realizing a closed feedback loop between agents and the data layers – where agents adapt their queries or quorum strategies based on cache or system feedback – remains an open problem.

## V. CONCLUSIONS

In this paper, we reimagine the data system as an adaptive collaborator rather than a passive executor in the face of dynamic agentic workloads. Traditional architectures, designed for static and predictable workloads, cannot accommodate the fluid, non-deterministic, and behavior-driven data access patterns exhibited by modern multi-agent systems. By introducing the concept of an Agent-centric Data Fabric, we propose a unifying framework that integrates attention-guided retrieval, semantic micro-caching, predictive prefetching and quorum-based serving to support dynamic agentic workloads.

This architecture establishes the foundation for context-aware, cost-sensitive, and cooperative data systems. Our exploration of open challenges underscores the need for new research in multimodal cost modeling, adaptive attention mechanisms, dynamic prefetching, semantic cache consistency, probabilistic quorum serving, and interoperable data protocols.

Looking ahead, realizing this vision will require joint advances in data management, distributed systems, and machine learning. Progress in these directions promises to redefine how future data systems interact with reasoning agents, thus shifting from static query processing toward behaviorally responsive data ecosystems that learn, adapt, and collaborate at scale.